\documentstyle[preprint,pra,aps,eqsecnum,amssymb,amsmath]{revtex}
\begin{document}
\preprint{}
\draft
%
%%%%%%%%%%%%%%%%%%%%%%%%%%%%%%%%% TITLE PAGE
%
\title{Renormalization Group and Probability Theory }
\author{G. Jona-Lasinio}
\address{Dipartimento di Fisica, Universit\`a "La Sapienza"\\
Piazzale A. Moro 2, 00185 ROMA, Italy}

%\date{\today}
\maketitle
%
%%%%%%%%%%%%%%%%%%%%%%%%%%%%%%%%% ABSTRACT
%
\begin{abstract}
The renormalization group has played an important role in the physics
of the second half of the twentieth century both as a conceptual and
a calculational tool. In particular it provided the key ideas for the construction 
of a qualitative and quantitative theory of the critical point 
in phase transitions and started a new era in statistical mechanics. 
Probability theory lies at the foundation
of this branch of physics and the renormalization group has an interesting probabilistic
interpretation as it was recognized in the middle seventies. 
This paper intends to provide a concise introduction to this aspect of the theory
of phase transitions which clarifies the deep statistical significance
of critical universality.        
\end{abstract}
%
%%%%%%%%%%%%%%%%%%%%%%%%%%%%%%%%% PACS NUMBERS
%
\pacs{ }
%
%%%%%%%%%%%%%%%%%%%%%%%%%%%%%%%%% PAPER CONTENT
%
\narrowtext
\tableofcontents
\section{Introduction}
The renormalization group (RG) is both a way of thinking and a
calculational tool which acquired its full maturity in connection 
with the theory of the critical point in phase transitions. 
The basic physical idea of the RG is that when we deal with systems 
with infinitely many
degrees of freedom, like thermodynamic systems, there are relatively
simple relationships between properties at different space scales
so that in many cases we are able to write down explicit exact
or approximate equations which allow us to study asymptotic
behaviour at very large scales.    

The first systematic use of probabilistic methods in statistical
mechanics was made by Khinchin who showed, using the
central limit theorem, that the Boltzmann distribution of the 
single-molecule energy in systems of weakly correlated
molecules is universal, that is independent of the form of the 
interaction, provided
it is of short range. In his well known book \cite{KH} he 
emphasizes that physicists had not fully appreciated the generality
of probabilistic methods so that, for example, they provided
a new derivation (usually heuristic) of the  Boltzmann law for every
type of interaction. Similar remarks apply to the first applications
of the RG in statistical mechanics. RG was 
introduced as a  tool to explain theoretically the universality
phenomena, the scaling laws, discovered experimentally near the
critical point of a phase transition. In the first period  RG
calculations used different formal devices, mostly borrowed from
quantum field theory, which gave good qualitative and quantitative 
results.
It was soon realized that a new class
of limit theorems in probability was involved. This class
referred to situations of strongly correlated variables to which
the central limit theorem does not apply, that is
situations opposite to those considered by Khinchin.
In fact, it was discovered that a critical point can be characterized   
by deviations from the central limit theorem.

Before providing a  description of the content of the
present paper we give a short account of the development
of RG ideas in statistical mechanics.

It is useful to distinguish two different conceptual approaches. 
The first use of the RG
in the study of critical phenomena \cite{DJ} was based on a Green's
function approach to statistical mechanics which paralleled 
quantum field theory. We recall Eq. (1) of \cite{DJ} 
\begin{equation}
{\cal G}(x, \{y_i\}, \alpha)= Z(t, \{y_i\}, \alpha) ~
{\cal G}(x/t, \{y_i/t\},
\alpha Z_V^{-1}(t, \{y_i\}, \alpha)Z^2(t, \{y_i\}, \alpha)),
\label{DJ}
\end{equation}
where ${\cal G}$ is a dimensionless two-point Green function depending
on a momentum variable $x$,  a set of physical
parameters $y_i$ and
the intensity of the interaction $\alpha$ (all dimensionless).
This is an
exact generalized scaling relation  which in the vicinity of the
critical point reduces to the phenomenological scaling due to the
disappearance of the irrelevant parameters. The scaling functions
$Z$ and $Z_V$ can be expressed in terms of the Green's functions themselves
via certain normalization conditions.
This equation provided a qualitative explanation of scaling 
and,  after the introduction of a non integer space dimension d and the use of 
$\epsilon=4-d$ as a perturbation parameter \cite{WF}, became
the basis for systematic quantitative calculations \cite{DC}, \cite{MI}, \cite{BLZ}.

The second approach started with the use on the part of Wilson
of a different  notion of RG \cite{W} that he had already 
introduced  in a different context, the fixed source meson theory
\cite{W65}, \cite{W70}, with no reference to critical phenomena. This
was akin to certain intuitive ideas of Kadanoff \cite{KA} about the
mechanism
of reduction of relevant degrees of freedom near the critical point.
Kadanoff's idea was that in the critical regime a thermodynamic
system, due to the strong correlations among the microscopic 
variables, behaves as if constituted by rigid blocks of arbitrary
size. In Wilson's approach in fact  the calculation of a statistical
sum consisted in a progressive  elimination of
the microscopic degrees of freedom to obtain the asymptotic large
scale properties of the system. 

Formally the Green's function and the Wilson
method were very different and in particular the first one
implied a true group structure while the second  was
a semigroup. Both gave exactly the same results and the problem
arose of clarifying the conceptual structures underlying
these methods. In fact many people were confused by this
situation and some thought that the two methods had little connection
with each other.
Actually the possibility of different RG transformations
equally effective in the study of critical properties could be  easily
understood using concepts from the theory of dynamical systems.
The critical point corresponds to a fixed point of these
transformations and the quantities of physical interest, i.e. the
critical indices, are connected with the hyperbolic behaviour
in its neighborhood, which is preserved if the different
transformations are related by a differentiable map \cite{JNO}.
Still the multiplicative structure of the Green's function RG
and the elimination of degrees of freedom typical of
Wilson's approach did not appear easy to reconcile. 
The formal connection was clarified in \cite{MANY} where
it was shown that to any
type of RG transformation one can associate a multiplicative
structure, a cocycle, and the characterizing feature of the
Green's function RG is that it is defined directly in terms of 
this structure. In the probabilistic setting the multiplicative
structure is related to the properties of conditional expectations
as discussed in \cite{CJL2} and illustrated in the present
paper. 
The relationship between the two approaches is an aspect that has not
been fully
appreciated in the literature and even an authoritative recent
exposition
of the RG history  seems to suggest
the existence of basic conceptual differences  \cite{FI}. In particular
Fisher discusses whether
equations 
like (\ref{DJ}) did anticipate Wilson and concludes
negatively. Different interpretations in the history of scientific
ideas 
are of course legitimate but after thirty years of applications of  RG
in critical phenomena and other fields  this conclusion does not appear
justified. In his 1970 paper
\cite{W70} on meson theories Wilson had emphasized
analogies of his approach with the Green function RG of  Gell-Mann and
Low even though a detailed comparison was not yet available. 
A balanced presentation
of the different RG approaches to critical phenomena  can be found in
the second edition of \cite{PP} of which, to my knowledge,
there is no English translation.

More difficult was to understand at a deeper level the
statistical nature of the critical universality. After 25 years
I still think that the language of probability provides
the clearest description of what is involved. The first hint
came from a study by Bleher and Sinai \cite{BS} of Dyson's
hierarchical models where they showed that at the critical point
the increase of fluctuations required a normalization different
from the square root of the number of variables in order to
obtain a non singular distribution for sums of spin variables (block
spin). 
This normalization factor is directly related to the rescaling parameter
of the fields in the RG.
The limit distribution could be either a Gaussian as in the
central limit theorem (CLT) or a different one which could
be calculated approximately. The next step consisted
in the recognition that new limit theorems for
random fields were involved \cite{J}, \cite{JG}, \cite{SI1}. 

The random fields appearing in these limit theorems have scaling
properties and some examples had already appeared in the
probabilistic literature. However these examples \cite{JCA} were not 
of a kind natural in statistical mechanics.
The new challenging problem posed by the theory
of phase transitions was the case of short range interactions
producing at the critical point long range correlations whose
scaling behaviour cannot be easily guessed from the microscopic
parameters. A general theory of such limit theorems is still
missing and so far rigorous progress has been obtained 
in situations which are not hierarchical but share with these
the fact that some form of scaling is introduced from the beginning.

The  main part of the present article will review the
connections 
of RG with limit theorems as they were understood in the decade
1975-85. The justification for presenting old material resides 
in the fact that these results are scattered in many different
publications very often with different perspectives. Here an effort
is made to present the probabilistic point of view in a synthetic
and coherent way.  
Section {\bf{X}} will  give an idea of more recent
work  trying to extract some general feature from the hard
technicalities which characterize it.

\section{A Renormalization Group derivation of the Central Limit
Theorem}
CLT asserts the following. Let $\xi_1, \xi_2,\ldots,\xi_n,\ldots$ be a
sequence of independent identically distributed (i.i.d.) random
variables with finite variance $\sigma^2={\mathbb{E}}
(\xi_i - {\mathbb{E}}(\xi_i))^2$,
where ${\mathbb{E}}$ means expectation with respect to their common distribution. Then
\begin{equation}
{{\sum_1^n (\xi_i - {\mathbb{E}}(\xi_i))}\over{\sigma n^{1/2}}} 
\stackrel{n \to \infty}{\longrightarrow} N(0,1)
\end{equation}
where the convergence is in law and $N(0,1)$ is the normal centered
distribution of variance $1$.
To visualize things consider the random variables $\xi_i$ as
discrete or continuous spins associated to the points of a
one dimensional lattice $\bf {Z}$ and introduce the block variables
$\zeta_n^1 = 2^{-{n/2}} \sum_1^{2^n} \xi_i$ and $\zeta_n^2 = 2^{-{n/2}}
\sum_{2^n+1}^{2^{n+1}} \xi_i$ 
Then
\begin{equation}
\zeta_{n+1} = {\frac{1} {\sqrt{2}}}  (\zeta_n^1 +  \zeta_n^2).
\end{equation}
Therefore we can write the recursive relation for the corresponding
distributions
\begin{equation}
p_{n+1}(x)=\sqrt{2}\int dy~ p_n(\sqrt{2}x-y)p_n(y) = ({\cal R} p_n)(x).
\label{R}
\end{equation}
The non linear transformation ${\cal R}$ is what we call a renormalization
transformation. Let us find its fixed points, i.e. the solutions of the
equation ${\cal R} p=p$. An easy calculation shows that the family
of Gaussians
\begin{equation}
p_{G,\sigma}(x)={{1}\over {\sqrt{2\pi \sigma^2}}} e^{-{{x^2}\over
{2\sigma^2}}}
\end{equation}
are fixed points. 
To prove the CLT we have to discuss the
conditions under which the iteration of ${\cal R}$ converges to a fixed 
point of variance $\sigma^2$. 
The standard analytical way is to use the Fourier  transform since ${\cal R}$
is a convolution. In view of the subsequent developments here
we shall illustrate the mechanism of convergence in the neighborhood of
a fixed point from the point of view
of nonlinear analysis. There are three conservation laws associated
with ${\cal R}$: normalization, centering and variance. In formulas
\begin{equation}
\int p_{n+1}(x)dx = \int p_{n}(x)dx ,
\end{equation}
\begin{equation}
\int xp_{n+1}(x)dx = \int xp_{n}(x)dx, 
\end{equation}
\begin{equation}
\int x^2p_{n+1}(x)dx = \int x^2p_{n}(x)dx .
\end{equation}
Therefore only distributions with variance $\sigma^2$ 
can converge
to a Gaussian $p_{G,\sigma}(x)$. 
We fix $\sigma=1$ and write $p_{G}$ for $p_{G,1}$.
Let us write the
initial distribution as a centered deformation of the Gaussian
with the same variance
\begin{equation}
p_{\eta}(x)=p_{G}(x)(1+\eta h(x))
\end{equation}
where $\eta$ is a parameter. The function $h(x)$ must satisfy
\begin{eqnarray}
\int p_G(x)h(x)dx=0,\\
\int p_G(x)xh(x)dx=0,\\
\int p_G(x)x^2h(x)dx=0.
\end{eqnarray}
Suppose now  $\eta$  small. In linear approximation we have
\begin{equation}
({\cal R} p_{\eta})= p_G (1+\eta({\cal L} h)) + {\cal O}(\eta^2),
\end{equation}
where $\cal L$ is the linear operator
\begin{equation}
({\cal L} h)(x)=2 \pi^{-1/2} \int dye^{-y^2}h(y+x2^{-1/2}).
\end{equation}
The eigenvalues  of $\cal L$ are
\begin{equation}
\lambda_k = 2^{1-k/2}
\end{equation}
and the eigenfunctions the Hermite polynomials. The three
conditions above on $h(x)$ can be read as the vanishing of its
projections on the first three Hermite polynomials.

The mechanism of convergence of the deformed distribution to
the normal law is now clear in linear approximation: if we develop
$h$ in Hermite polynomials only terms with $k > 2$ will appear so that
upon iteration of the RG transformation they will contract
to zero exponentially as the corresponding eigenvalues are $< 1$.

To complete the proof one must show that the non linear terms
do not alter the conclusion. This is less elementary and will not be
pursued here. 

A terminological remark. The Gaussian is an example of what is
called in probability theory a stable distribution. These are
distributions which are fixed points of convolution equations and, 
with the exception of the Gaussian, have infinite variance.

\section{Hierarchical Models}
Suppose now that the $\xi_i$ are not independent. A case which has
played a very important role in the development of the RG
theory of  critical phenomena is that of hierarchical models.
To keep the notation close to that of the previous section we 
write the recursion relation connecting the distribution at level $n$
to that at level $n+1$
\begin{equation}
p_{n+1}(x)=({\hat{{\cal R}}}p_n)(x)=g_n(x^2)({\cal R}p_n)(x),
\label{H}
\end{equation}
where $g_n(x^2)$ is a sequence of positive increasing functions and ${\cal R}$
has the same meaning as in the previous section. It is clear that
such a dependence tends to favor large values of the block variable
$x$ and therefore values of the $\xi_i$'s of the same sign.
We call this a ferromagnetic dependence. We make the following choice 
$g_n(x^2)=L_ne^{\beta (c/2)^nx^2}$, where the constant
$L_n$ is determined by the normalization condition. This type of
recursion arises from the following Gibbs distribution
\begin{equation}
d\mu=Z_n^{-1}e^{-\beta H_n(x_1,\ldots,x_{2^n})} \prod_1^{2^n}dp_0(x_i),
\end{equation}
where $H_n$ has the following hierarchical structure
\begin{equation}
H_n(x_1,\ldots,x_{2^n}) = H_{n-1}(x_1,\ldots,x_{2^{n-1}}) + 
H_{n-1}(x_{2^{n-1}+1},\ldots,x_{2^n}) - c^n 
\left( \sum_1^{2^n}\frac{x_i}{2} \right)^2   ,
\end{equation}
$H_0 = 0$ and $p_0(x)$ is a single spin distribution which
characterizes the model. The constant $c$ satisfies
$1<c<2$. For $c<1$ the model is trivial while for $c>2$
becomes thermodynamically unstable. 

To understand what happens in the case of dependent variables
let us consider the hierarchical model defined by a Gaussian single
spin
distribution where the iteration can be performed exactly,
that is
\begin{eqnarray}
p_0(x) &=& {(2\pi)}^{-1/2}e^{-x^2/2}, \\
({\hat{{\cal R}}}^np_0)(x) &=& (2\pi \sigma_n^2)^{-1/2} e^{-x^2/2 \sigma_n^2}, \\
\sigma_n^2 &=& \left( 
1-2\beta \sum_{1}^n (c/2)^k \right)^{-1}.
\end{eqnarray}
We see that the conservation of variance does not hold anymore
under the transformation $\hat {\cal R}$ and in fact the variance increases
at each iteration and is $\beta$ dependent. When $n$ tends
to infinity the iteration converges to  the distribution 
\begin{equation}
p(x)={(2\pi)}^{-1/2}{ (1-2\beta c/(2-c))}^{1/2}e^{-(1-2\beta c/(2-c))x^2/2}
\end{equation}
provided $\beta < \beta_{cr}=1/c-1/2$, that is the CLT holds
if the temperature is sufficiently large. At $\beta = \beta_{cr}$
the variance of the limit distribution explodes which means that
the fluctuations increase faster than ${\cal O}(2^{n/2})$. We try a new
normalization of the block variable and consider $\sum_1^{2^n} \xi_i/2^nc^{-n/2}$. The recursion for
the distribution of this variable is
\begin{equation}
p_{n+1}(x)=L_ne^{\beta x^2}\int dyp_n(2/c^{1/2}x - y)p_n(y).
\label{RGH}
\end{equation}
We now follow the same pattern as in the previous section:
calculate the fixed points and see whether they admit a stable
manifold or, in probabilistic language, a domain of attraction.
The fixed points are the solutions of the equation
\begin{equation}
p(x)=Le^{\beta x^2}\int dyp(2/c^{1/2}x - y)p(y)
\label{FP}
\end{equation}
with $L$ determined by normalization. A Gaussian solution is
easily found
\begin{equation}
p_G(x)={(a_0/\pi)}^{1/2}e^{-a_0x^2}
\label{GA}
\end{equation}
with $a_0=c\beta /(2-c)$.
We shall again discuss  stability in linear approximation by
considering a centered deformation of (\ref{GA})
$p_{\eta}(x)=p_G(x)(1+\eta h(x))$. For small $\eta$ 
\begin{equation}
({\hat {\cal R}} p_{\eta})= p_G (1+\eta({\hat{\cal L}} h)) + {\cal O}(\eta^2).
\end{equation}
where ${\hat{\cal L}}$ is the linear operator
\begin{equation}
({\hat{{\cal L}}}h)(x)=\int dye^{-2a_0y^2}(h((x+y)/c^{1/2}) +
h((x-y)/c^{1/2}))
\end{equation}
The corresponding eigenvalues for even eigenfunctions are 
$\lambda_{2k}=2/c^k$, with $k=0,1,2,3,\ldots$, and the eigenfunctions
are rescaled Hermite polynomials of even degree.

Here $h(x)$ must be considered as the effect of many iterations starting 
from
some initial distribution which characterizes the model and is
therefore dependent on $\beta$. We now see that  for $2>c>2^{1/2}$
the eigenvalues $\lambda_0$ and $\lambda_1$ are $>1$. The projection of
$h$ over the constants vanishes due to the normalization
condition so that for the iteration to converge we have to
impose the vanishing of the projection of $h$ on the second
Hermite polynomial. In view of the previous remark this will
select  a special value $\beta_{cr}$, the critical temperature for the
model considered. In conclusion, for  $2>c>2^{1/2}$ and $\beta = \beta_{cr}$ the fixed point
(\ref{GA}) has a non empty domain of attraction.

When $c<2^{1/2}$ the Gaussian fixed point becomes unstable and we must
investigate about the existence of other fixed points.
Bifurcation theory tells us that most likely there is an exchange
of stability between two fixed points and we should look for
the new one in the direction which has just become unstable.
For $c<2^{1/2}$ we have $\lambda_4>1$ so the instability
is in the direction $H_4$, the Hermite polynomial of fourth order.
Define $\epsilon = 2^{1/2}-c$ and look, for $\epsilon$ small, for a
solution of (\ref{FP}) of the form 
\begin{equation}
p_{NG}(x)=p_G(x)(1-\epsilon aH_4(\gamma x)) + {\cal O}(\epsilon^2)
\approx e^{-r^*(\epsilon)x^2/2 - u^*(\epsilon)x^4/4},
\label{NG}
\end{equation}
where $r^*$ and $u^*$ are the fixed point couplings.
The analysis of this case is considerably more complicated and gives
the following results: the linearization of the RG transformation
around (\ref{NG}) has only one unstable direction
so that by requiring the vanishing of an appropriate projection
along this direction we obtain a non empty domain of attraction
for some $\beta_{cr}$ \cite{BS1}, \cite{CEK}.  
 
\section{Eigenvalues of the linearized RG and critical indices}
We illustrate the interpretation of the eigenvalues of the linearized RG
at a fixed point in the context of
hierarchical models, which is especially simple.
Notations are as in the previous section. Consider for definiteness the 
region $2^{1/3}<c<2^{1/2}$ so that the Gaussian fixed point  
is unstable, but in its
neighborhood there exists a  non trivial non Gaussian fixed point
of the form (\ref{NG}) with a non empty domain of attraction.
Suppose now that we start the iteration
of the RG from some initial distribution which is close to it but
not in its domain of attraction. For example we may consider
a distribution $p(x, \beta)$ of the form (\ref{NG}) with parameters 
$r, u$
slightly different from $r^*, u^*$ which are the values
taken at the critical temperature $\beta_{cr}$. Application of the RG transformation
will eventually drive this distribution away from the fixed
point due to the presence of an unstable direction. 
However we can "renormalize" our parameters $r, u$ by compensating
the instability at each iteration and find a sequence of parameters
$r_n, u_n$ such that when $n$ tends to infinity we have a sequence of
distributions $p_n(x, \beta_n)$ approaching a definite limit.
Since we have assumed that the parameters $r_n, u_n$ are close
to the fixed point values, 	the renormalized parameters can
be simply expressed in terms of rescalings determined by the
eigenvalues of a linear operator  analogous to ${\hat{\cal L}}$ introduced
in the previous section in connection with the Gaussian fixed point.    
This can be seen as follows. Let us write our initial distribution
as a deformation of (\ref{NG})
\begin{equation}
p(x, \beta)=p_{NG}(x, \beta_{cr})(1 + \eta h(x, \beta)).
\end{equation}
If we develop $h$ in terms of eigenfunctions of the linearized RG at (\ref{NG}),
the iteration of the RG transformation will multiply the projections of $h$ along
these eigenfunctions by powers of the corresponding eigenvalues. The explosion
in the unstable direction can then be controlled by rescaling at each step
the projection in this direction with a factor proportional to the inverse
eigenvalue.  

Let us define the susceptibility of a block of $2^n$ spins
\begin{equation}
\chi_n(\beta) = {\frac {1}{2^n}}{\mathbb{E}}\left[ \left(\sum_1^{2^n}\xi_i\right)^2\right]
\label{SU}
\end{equation}
This quantity can be easily expressed in terms of the distribution
$p_n(x, \beta)$ of the block with the critical normalization 
$2^nc^{-n/2}$ 
\begin{equation}
\chi_n(\beta) = (2/c)^n\int dxx^2 p_n(x, \beta).
\end{equation}
As $n\rightarrow \infty$, $\chi_ n$ diverges if $\beta = \beta_{cr}$.

 To calculate the susceptibility critical index as $\beta$ approaches
the critical value we assume that the two limits $n\rightarrow\infty$
and $\beta \rightarrow\beta_{cr}$ can be interchanged so that they can be
calculated over subsequences. We want to compute
\begin{equation}
\nu_{\chi}=\lim_{n\rightarrow\infty} {\frac {\log \chi_n(\beta_n)} 
{\log |\beta_n - \beta_{cr}|}}
= \lim_{n\rightarrow\infty} {\frac {-n\log (c/2) + \log \int dy ~y^2 p_n(y, \beta_n)}
{\log |\beta_n - \beta_{cr}|}}.
\end{equation}
If we now choose $|\beta_n - \beta_{cr}|\approx \lambda^{-n}$, where
$\lambda$ is the eigenvalue corresponding to the unstable direction
of the RG linearized at the fixed point, the integral appearing in
this formula will be almost constant and we obtain
\begin{equation}
\nu_{\chi}={\frac {\log (c/2)} {\log \lambda}}.
\end{equation}
Similar calculations can be done for other thermodynamical quantities
like the free energy or the magnetization.

We can summarize the situation as follows: given a model
defined by an initial distribution $p_0$, for $\beta<\beta_{cr}$
we expect the CLT to hold. For $\beta=\beta_{cr}$ by properly
normalizing the block variables we have new limit theorems
where the limit law has a domain of attraction which is a non trivial
submanifold in the space of probability distributions called the
critical manifold. If we start from a distribution which is not
in the domain of attraction of a given fixed point, but not too far
from it, it can still be driven  to a regular limit by rescaling at
each step its coefficients   in a way dictated by the fixed point.
This defines the so called scaling limits of the theory associated
to a given fixed point. 

For further reading see \cite{CEK}, \cite{CG}, \cite{GK}.

\section{Self Similar Random Fields}
The notion of self similar random field was introduced informally in\
\cite{J} and rigorously in \cite{JG} and independently in \cite{SI1}.
It was then developed more systematically in \cite{DO1} and \cite{DO2}.
The idea was to construct a proper mathematical setting for the
notion of RG {\it a la} Kadanoff-Wilson. This led to a generalization
of limit theorems for random fields to the situation in which the
variables are strongly correlated.

Let ${\bf {Z}}^d$ be a lattice in $d$-dimensional space and $j$ a generic
point of ${\bf {Z}}^d$, $j=(j_1, j_2, ...,j_d)$ with integer coordinates
$j_i$. We associate to each site a centered random variable $\xi_j$ and
define
a new random field
\begin{equation}
\xi_j^n=({\cal R}_{\alpha,n}\xi)_j=n^{-d\alpha/2}\sum_{s\in V_j^n} \xi_s,
\label{RF}
\end{equation}
where
\begin{equation}
V_j^n=\{s: j_kn - n/2 < s_k \leq j_kn + n/2 \}
\end{equation}
and $1\leq\alpha <2$. The transformation (\ref{RF}) induces a
transformation on probability measures according to
\begin {equation}
({\cal R}^*_{\alpha,n}\mu)(A)={\mu'}(A)=\mu({\cal R}^{-1}_{\alpha,n}A),
\label{RMU}
\end{equation}
where $A$ is a measurable set and ${\cal R}^*_{\alpha,n}$ has the semigroup
property
\begin{equation}
{\cal R}^*_{\alpha,n_1}{\cal R}^*_{\alpha,n_2}={\cal R}^*_{\alpha,n_1+n_2}.
\end{equation}
A measure $\mu$ will be called  self similar if
\begin{equation}
{\cal R}^*_{\alpha,n}\mu=\mu
\label{ST}
\end{equation}
and the corresponding field will be called a  self similar random
field.
We briefly discuss the choice of the parameter $\alpha$.
It is natural to take $1\leq\alpha<2$. In fact $\alpha=2$ corresponds
to the law of large numbers so that  
the block variable (\ref{RF}) will tend for large $n$ to zero in probability.
The case $\alpha >1$
means that we are considering random systems which fluctuate 
more than a collection of independent variables and $\alpha=1$ 
corresponds to the CLT. Mathematically the lower bound is not
natural but it becomes so when we restrict ourselves to the 
consideration of ferromagnetic-like systems.

A general theory of self similar random fields so far does not exist
and presumably is very difficult. However Gaussian fields are
completely specified by their correlation function and self similar
Gaussian fields can be constructed explicitly \cite{SI1}, \cite{SI2}. 
It is easier if we represent the correlation function in terms of
its Fourier transform
\begin{equation}
{\mathbb{E}}(\xi_i\xi_j) = \int_{-\pi}^\pi \prod_1^d d\lambda_k \rho (\lambda_1,\ldots,\lambda_d)e^{i\sum_k \lambda_k {(i-j)}_k}.
\label{FT}
\end{equation}   
The prescription to construct $\rho$ in such a way that the
corresponding Gaussian field satisfies (\ref{ST}) is as follows.
Take a positive homogeneous function $f(\lambda_1,...,\lambda_d)$ with
homogeneity exponent $d(1+\alpha)$, that is
\begin{equation}
f(c\lambda_1,...,c\lambda_d)=c^{d(1+\alpha)}f(\lambda_1,...,\lambda_d)
\end{equation}.
Next we construct a periodic function $g(\lambda_1,...,\lambda_d)$ by
taking an average over the lattice $Z^d$
\begin{equation}
g(\lambda_1,...,\lambda_d)=\sum_{i_k} {{1}\over {f(\lambda_1 +
i_1,\ldots,\lambda_d + i_d)}}.
\end{equation}
If we take now
\begin{equation}
\rho (\lambda_1, \ldots,\lambda_d)=\prod_i {|1-e^{i\lambda_i}|}^2
g(\lambda_1,\ldots,\lambda_d),
\end{equation}
it is not difficult to see that the corresponding Gaussian measure
satisfies (\ref{ST}). The periodicity of $\rho$ insures translational
invariance.

For $d=1$ there is only one, apart from a multiplicative constant,
homogeneous function and one can show that the above construction
exhausts all possible Gaussian self similar distributions. For $d>1$ it
is not known whether a similar
conclusion holds.

From this point on one can follow in the discussion the same pattern as
for hierarchical models and investigate the stability
of the Gaussian fixed points $P_G$ of (\ref{ST}).  Consider a
deformation $P_G(1+h)$ and the action of ${\cal R}^*_{\alpha,n}$
on this distribution. It is easily seen that
\begin{equation}
{\cal R}^*_{\alpha,n} P_G h =  {\mathbb{E}}(h|\{\xi_j^n\}) {\cal R}^*_{\alpha,n} P_G =
{\mathbb{E}}(h|\{\xi_j^n\}) P_G(\{\xi_j^n\}).
\label{LR}
\end{equation}
The conditional expectation on the right hand side of (\ref{LR})
will be called the linearization of the RG at the fixed point $P_G$.
To proceed further in the study of the stability we have to
find its eigenvectors and eigenvalues. These have been calculated
by Sinai. The eigenvectors are appropriate infinite dimensional
generalizations of Hermite polynomials $H_k$ which are described in full
detail in \cite{SI2}. They satisfy the eigenvalue equation
\begin{equation}
{\mathbb{E}}(H_k|\{\xi_j^n\}) = n^{[k(\alpha /2 - 1) + 1]d}H_k(\{\xi_j^n\}).
\label{FPE}
\end{equation} 
We see immediately that  $H_2$ is always unstable while $H_4$ becomes
unstable when $\alpha$ crosses from below the value $3/2$.  By introducing
the parameter $\epsilon = \alpha - 3/2$, in principle one can construct, 
as in the hierarchical case, a non Gaussian fixed point. 
The formal construction is explained in
Sinai's book \cite{SI2} where one can find also an exhaustive discussion of the
questions, mostly unsolved, arising in this connection.
A different construction of a non Gaussian fixed point, 
for $d=4$ has been made recently by Brydges, Dimock
and Hurd. This will be briefly discussed in section X. 

\section{Some Properties  of Self Similar Random Fields}
We have already  characterized the critical point as a
situation of strongly dependent random variables,
in which the CLT fails. We want to give here a characterization
which refers to the random field globally. Consider in the
product space of the variables $\xi_i$ the cylinder sets, that is
the sets of the form
\begin{equation}
\{\xi_{i_1}\in A_1,\ldots,\xi_{i_n}\in A_n\},
\end{equation}
with $i_1,\ldots,i_n \in  \Lambda$, $\Lambda$ being an arbitrary
finite region in ${\bf {Z}}^d$ and the $A_i$ measurable sets in the
space of the variables $\xi_i$. We denote with $\Sigma_{\Lambda}$
the $\sigma$-algebra generated by such sets. We say that the variables
$\xi_i$ are {\sl{weakly dependent}} or that they are a {\sl
{strong
mixing}} random field if the following holds. Given two finite regions
$\Lambda_1$ and $\Lambda_2$ separated by a distance
\begin{equation}
d(\Lambda_1, \Lambda_2)=\min_{i \in \Lambda_1, j\in \Lambda_2}
|i-j|,
\end{equation}
where $|i-j|$ is for example the Euclidean distance, define
\begin{equation}
\tau(\Lambda_1, \Lambda_2)=\sup_{A\in\Sigma_{\Lambda_1},
B\in\Sigma_{\Lambda_2}}|\mu(A\cap B)-\mu (A)\mu (B)|.
\end{equation}
Then $\tau(\Lambda_1, \Lambda_2) \to 0$ when
$d(\Lambda_1, \Lambda_2) \to \infty$. 

Intuitively the strong mixing idea is that one cannot compensate
for the weakening of the dependence of the variables due to an
increase of their space distance, by increasing the size of the sets.

This situation is typical when one has exponential decay of
correlations. This has been proved for a wide class   
of random fields including ferromagnetic non critical spin
systems \cite{HN}.

The situation is entirely different at the critical point where one
expects the correlations  to decay as an inverse power of the distance.
In this connection the following result has been proved in \cite{CJL}:
a ferromagnetic translational invariant
system with pair interactions with correlation function
\begin{equation}
C(i)={\mathbb{E}}(\xi_0\xi_i) - {\mathbb{E}}(\xi_0){\mathbb{E}}(\xi_i)
\end{equation}
such that
\begin{equation}
\lim_{L\rightarrow \infty} {{\sum_{L(s_k-1)\leq i_k
<L(s_k+1)}C(i)}\over{\sum_{0\leq i_k < L} C(i)}} \neq 0
\end{equation}
for arbitrary $s_k$, does not satisfy the strong mixing
condition.

This theorem implies in particular that a critical 2-dimensional 
Ising model violates strong mixing. Therefore violation of strong
mixing seems to provide a reasonable characterization of the type
of strong dependence encountered in critical phenomena. On the other
end, under very general conditions, if strong mixing holds
the one-block distribution satisfies the CLT \cite{IL}.

An interesting question is whether we can describe the structure
of the limit one-block  distributions that  can appear at the critical
point beside the Gaussian. It was shown in \cite{CJL}, building
on previous results by Newman,  that for
ferromagnetic systems the Fourier transform (characteristic function in
probabilistic language) of the limit distribution must
be of the form
\begin{equation}
{\mathbb{E}}(e^{it\xi})=e^{-bt^2} \prod_j (1-t^2/\alpha_j^2)
\end{equation}
with$\sum_j 1/\alpha_j^2 < \infty$. In the probabilistic literature
these distributions are called the $D$-class \cite{LO}. The Gaussian
is the only infinitely divisible distribution belonging to this class. 

\section{Multiplicative Structure}
In this section we show that there is a natural multiplicative
structure associated with transformations on probability
distributions like those induced by the RG. This multiplicative
structure is related to the properties of conditional expectations.
We use the notations of section V. Suppose we wish to evaluate
the conditional expectation
\begin{equation}
{\mathbb{E}}(h|\{\xi_j^n\}),
\end{equation}
where the collection of block variables $\xi_j^n$ indexed by $j$
is given a fixed value. Here $h$ is a function of the individual spins
$\xi_i$. It is an elementary property of conditional expectations 
that
\begin{equation}
{\mathbb{E}}({\mathbb{E}}(h|\{\xi_j^n\})|\{\xi_j^{nm}\}) = {\mathbb{E}}(h|\{\xi_j^{nm}\}).
\label{EX}
\end{equation}
Let $P$ be the probability distribution of the $\xi_i$ 
and ${\cal R}^*_{\alpha,n}$ the distribution obtained by applying the RG transformation,
that is the distribution of the block variables $\xi^n_j$. By
specifying in (\ref{EX}) the distribution with respect to which
expectations are taken we can rewrite it as
\begin{equation}
{\mathbb{E}}_{{\cal R}^*_{\alpha,n}P}({\mathbb{E}}_P(h|\{\xi_j^n\})|\{\xi_j^{nm}\}) =
{\mathbb{E}}_P(h|\{\xi_j^{nm}\}).
\label{EX1}
\end{equation}
This is the basic equation of this section and we want to work out
its consequences. For this purpose we generalize the eigenvalue
equation (\ref{FPE}) to the case in which the probability distribution
is not a fixed point of the RG. In analogy with the theory of
dynamical systems we interpret the conditional expectation as
a linear transformation  from the linear space tangent to $P$ to
the linear space tangent to ${\cal R}^*_{\alpha,n}P$ and we assume that
in each of these spaces there is a basis of vectors $H^{P}_k$,
$H^{{\cal R}^*_{\alpha,n}P}_k$  connected by
the following generalized eigenvalue equation \cite{OS}
\begin{equation}
{\mathbb{E}}_P(H^P_k|\{\xi_j^n\}) = \lambda_k(n, P) H^{{\cal R}^*_{\alpha,n}P}_k(\{\xi_j^n\}).
\label{GEE}
\end{equation}
Equation (\ref{EX1}) implies that the $\lambda$'s must satisfy
the relationship
\begin{equation}
\lambda_k(m, {\cal R}^*_{\alpha,n} P)\lambda_k(n, P) = \lambda_k(mn, P).
\label{J}
\end{equation}   
From (\ref{GEE}) and (\ref{J}) we find that the $\lambda_k$ are 
given by the following expectations
\begin{equation}
\lambda_k(n, P) = {\mathbb{E}}({\bar{H}}^{{\cal R}^*_{\alpha,n}P}_k(\{\xi_j^n\})H^P_k(\{\xi_j\})),
\end{equation}
where ${\bar{H}}_k^P$ are dual to $H_k^P$ according to the
orthogonality relation $\int {\bar{H}}_k^P H_j^P dP = \delta_{kj}$.  
The $\lambda_k$ are therefore special correlation functions. 
The similarity
between  equation (\ref{J}) and (\ref{DJ}) is then obvious. The Green's function RG
corresponds to a very simple transformation on the probability distribution
such that its form is unchanged and only the values 
of its parameters are modified.

\section{RG and Effective Potentials}
In this section we want to illustrate a connection between 
RG and the theory
of large deviations \cite{JF}. By large deviations we
mean fluctuations with respect to the law of large numbers,
e.g., fluctuations of the magnetization in a large but finite
volume. In view of the connection of RG with limit theorems
our discussion will parallel, actually generalize, some well
known facts in the theory of sums of independent random
variables. This will lead to a probabilistic interpretation
of a widely used concept in physics, the effective potential,
and will clarify its relationship with the effective Hamiltonian
in RG theory. We continue with our model system of continuous
spins $\xi_i$ indexed by the sites of a lattice in $d$ dimensions
and try to estimate the probability that the magnetization in a
volume $\Lambda$ be larger than zero at some temperature
above criticality. From the exponential Chebysheff 
inequality we have for $\theta > 0, x > 0$
\begin{equation}
P\left(\sum_{i\in \Lambda} \xi_i/|\Lambda| \geq x\right) \leq
e^{-|\Lambda|\theta x} {\mathbb{E}}(e^{\theta 
\sum_{i\in \Lambda} \xi_i}) 
\leq e^{-|\Lambda|\Gamma (|\Lambda|, x)},
\end{equation}
where
\begin{equation}
\Gamma (|\Lambda|, x)=\sup_{\theta > 0}(\theta x - 
{\frac{1}{|\Lambda|}} \log  {\mathbb{E}}(e^{\theta \sum_{i\in \Lambda} \xi_i}))
\label{EP}
\end{equation}
is the Legendre transform of 
${\frac{1}{|\Lambda|}} \log  {\mathbb{E}}(e^{\theta \sum_{i\in \Lambda} \xi_i})$. 
With some more work one can
establish also a lower bound 
\begin{equation}
P\left(\sum_{i\in \Lambda} \xi_i/|\Lambda| \geq x\right) \geq
e^{-|\Lambda|(\Gamma (|\Lambda|, x) + \alpha (|\Lambda|) + \delta)},
\end{equation}
with $\alpha \rightarrow 0$ for $\Lambda \rightarrow \infty$,
and $\delta >0$ arbitrarily small. We then conclude
\begin{equation}
-\lim_{\Lambda \rightarrow \infty}{\frac{1}{|\Lambda|}}\log 
P\left(\sum_{i\in \Lambda}
\xi_i/|\Lambda| \geq x\right) =\lim_{\Lambda \rightarrow \infty} \Gamma
(|\Lambda|, x) =V_{eff}(x),
\end{equation}
where $V_{eff}(x)$ is known in the physical literature as the effective
potential. An important remark. While $\Gamma (|\Lambda|, x)$,
being the Legendre transform of a convex function, is always
convex for any $\Lambda$, this is not the case with
$-{\frac{1}{|\Lambda|}}\log P(\sum_{i\in \Lambda} \xi_i/|\Lambda| \geq x) =
V(|\Lambda|, x)$ for finite $\Lambda$ and one has to be careful in interpreting
for example results from numerical simulations.

To understand the connection with the RG it is convenient to 
consider first the case of independent random variables, that is
the situation considered in section II. A classical problem in limit
theorems for independent variables is the estimate of the corrections
to the CLT when the argument of the limit distribution
increases with $n$. A well known result in this domain is the
following \cite{IL}: suppose we want to estimate 
$P(\sum_1^n\xi_i/n^{1/2}\geq x)$ when $x=o(n^{1/2})$. Then 
\begin{equation}
P\left(\sum_1^n\xi_i/n^{1/2}\geq x\right)\approx e^{-n\sum_2^s \Gamma_k
(x n^{-1/2})^k}
\label{CLTC}
\end{equation}
for $n\rightarrow \infty$ and $\lim_{n\rightarrow \infty}
x^{s+1}n^{-(s-1)/2} = 0$. The function $\Gamma(z)=\sum_2^{\infty}\Gamma_k z^k$ is
the Legendre transform
of $\log {\mathbb{E}}(e^{\theta \xi_i})$. The sign $\approx$ has to be understood
as logarithmic dominance. If $x={\cal O}(n^{1/2})$, the
whole function $\Gamma$ contributes and we are back to
the large deviation estimate at the beginning of this section.

We expect a result like (\ref{CLTC}) to hold for the one-block
distribution  in the case of dependent variables as in statistical 
mechanics away from the critical point. We then see that the
coefficients of an expansion of $\Gamma(|\Lambda|, x)$ in
powers of $x$ determine the corrections to the CLT for the
one-block distribution. More interesting is the situation at the
critical point. Suppose first that the one block limit distribution
is Gaussian  but the normalization is anomalous as it is the case
in hierarchical models for a range of values of the parameter $c$.
Instead of (\ref{CLTC}) we expect an estimate of the form
\begin{equation}
P\left(\sum_{i\in \Lambda}\xi_i/{|\Lambda|}^{\rho}\geq x\right) \approx
e^{-|\Lambda|\sum_2^s \Gamma_k(|\Lambda|)
(x/{|\Lambda|}^{1-\rho})^k},
\end{equation}
with $\rho >1/2$ and $\lim_{|\Lambda|\rightarrow \infty} x^{s+1}
/{|\Lambda|}^{(1-\rho)s-\rho}=0$. We see that for the quadratic
term to survive the coefficient $\Gamma_2$ must vanish when
$|\Lambda| \rightarrow \infty$ as ${|\Lambda|}^{1-2\rho}$. 
If the one-block limit distribution is not Gaussian we can establish
a general relationship between its logarithm and the effective potential. Let us
write $-\log P = V_{RG}$ where $P$ is the limit
distribution. We now rewrite the large deviation estimate in the
following way
\begin{equation}
P\left(\sum_{i\in \Lambda} \xi_i/{|\Lambda|}^\rho \geq
x{|\Lambda|}^{1-\rho}\right) \approx
e^{-|\Lambda|\Gamma (|\Lambda|, x)}.
\end{equation}
Scale $x\rightarrow x/{|\Lambda|}^{1-\rho}$. Then
we obtain
\begin{equation}
V_{RG}(x) = \lim_{|\Lambda|\rightarrow \infty}
|\Lambda|\Gamma(|\Lambda|, x/{|\Lambda|}^{1-\rho}).
\end{equation}
Therefore $\Gamma (|\Lambda|, x)$ determines in different
limits either $V_{eff}$ or $V_{RG}$. The discussion in the
present section can be made rigorous in the case of
hierarchical models.

\section{Coexistence of Phases in Hierarchical Models}
In the case of hierarchical models the RG recursion relation
for the one-block probability distribution can be easily rewritten 
as a recursion for the quantity $V(|\Lambda|, x)$ introduced
in the previous section which coincides with the effective potential
in the limit $|\Lambda| \rightarrow \infty$. In fact if we normalize
the block-spin with its volume, that is consider the mean
magnetization, a simple calculation gives
the following iteration for the corresponding probability
distribution $\pi_n(x)$:
\begin{equation}
\pi_n(x)=L_ne^{\beta c^nx^2}\int dx' \pi_{n-1}(2x-x')\pi_{n-1}(x').
\end{equation}
Taking the logarithm and dividing by the number of spins $2^n$,
we obtain
\begin{equation}
V_n(x)=-1/2^n \log L_n - \beta {(c/2)}^n x^2 - 
1/2^n \log \int dx' e^{-2^{n-1}(V_{n-1}(2x-x') + V_{n-1}(x'))}.
\label{EPI}
\end{equation} 
To illustrate the difference between (\ref{EPI}) and (\ref{RGH})
let us consider again the simple case in which the model is defined
by a Gaussian single spin distribution. The iteration of (\ref{EPI})
gives for small $\beta$
\begin{equation}
V_n(x)=1/2 \left[1 - \beta \sum_0^n (c/2)^k\right] x^2 + \nu_n
\end{equation}
where $\nu_n$ tends to zero when $n \rightarrow \infty$.
In this limit then
\begin{equation}
V_{eff}=1/2[1 - 2c\beta /(2-c)] x^2.
\end{equation}
The critical temperature is defined as the value $\beta_{cr}$ for
which the coefficient of $x^2$ vanishes and coincides with that
found in section {\bf  III}. On the other hand in that section it was the
only temperature for which the recursion (\ref{RGH}) converges to the
Gaussian fixed point, i.e. the only temperature for which
the following difference between two diverging expressions converges
\begin{equation}
2\beta \sum_0^n {(2/c)}^k - {(2/c)}^{n+1}.
\end{equation}

We want to apply now (\ref{EPI}) to the study of the magnetization
in the phase coexistence region for a general hierarchical model
\cite{BL}, \cite{JHP}. The
problem we want to discuss
is the following. In the hierarchical case at level $n$ we have blocks,
containing each $2^{n-1}$ spins, interacting in pairs
through the Hamiltonian
\begin{equation}
c^n{(\zeta_{n-1}^1 + \zeta_{n-1}^2)}^2/4 = c^n (\zeta_n)^2
\end{equation}
where  $\zeta_{n-1}^1, \zeta_{n-1}^2, \zeta_n$ are mean
magnetizations. Suppose now that $\zeta_n$ is assigned the value
$\alpha M$, $0<\alpha <1$, $M$ being the spontaneous magnetization
corresponding to the temperature $\beta$.
We want to calculate the conditional distribution of $\zeta_{n-1}$
given $\zeta_n$, for large $n$. The remarkable result is that
one of the quantities $\zeta_{n-1}^1$ or $\zeta_{n-1}^2$
with probability close to $1$ is equal to the full magnetization $M$.

To compute the desired distribution we have to estimate $\pi_n(x)$
or, what is the same, $V_n(x)$ for large $n$. From (\ref{EPI}) we
expect asymptotically 
\begin{equation}
V_n(x) = V_{eff}(x) + {(c/2)}^n Y(x) + \ldots 
\end{equation}
Since in the phase coexistence region $V_{eff}(x)$ is flat, i.e.
constant, the whole $x$ dependence is given by $Y(x)$. In 
order to compute this function we perform a subtraction
and consider $V_n(x) - V_n(x_0)$ choosing $x_0$ in the 
flat region of $V_{eff}(x)$. From (\ref{EPI}) it is easily seen
that the quantity 
\begin{equation}
\Delta_n(x) = {(2/c)}^n(V_n(x) - V_n(x_0))
\end{equation}
satisfies a recursion of the form
\begin{equation}
\Delta_n(x) = - c^{-n} \log A_n - \beta x^2 
- c^{-n} \log \int dx' e^{-c^{n-1}(\Delta_{n-1}(x+x') +
\Delta_{n-1}(x-x'))},
\label{DELTA}
\end{equation}
where $A_n$ is determined by the condition $\Delta_n(x_0) = 0$. 
Let us choose $x_0 = M$. By symmetry $\Delta_n(\pm M) = 0$.
For $0\leq x \leq M$ and large $n$ the main contribution to the 
integral on the right hand side of (\ref{DELTA}) comes from
the region $x\pm x' \approx M$, while for $-M \leq x \leq 0$
from the region $x\pm x' \approx -M$. We can write therefore the
approximate recursion equations
\begin{equation}
\Delta_n(x) = \beta(M^2-x^2) + c^{-1}\Delta_{n-1}(2x\mp M),
\end{equation}
where the $\mp$ in the second term on the right corresponds
to $0\leq x \leq M$ or $-M\leq x\leq 0$. This type of equations
has been rigorously studied by Bleher \cite{BL} and the asymptotic
solutions show a complicated fractal structure.

The conditional probability of interest to us is
\begin{equation}
P(|\zeta_{n-1} - M| < \epsilon M| \zeta_n = \alpha M) =
\frac
{\int_{|x'-M| < \epsilon M} dx' e^{-c^{n-1}(\Delta_{n-1}(x') + 
\Delta_{n-1}(2\alpha M - x'))} }
{\int_\infty^\infty
 dx' e^{-c^{n-1}(\Delta_{n-1}(x') + \Delta_{n-1}(2\alpha M - x'))} }
\end{equation}
Since  the main contribution  to the integral
in the denominator comes from the same region appearing
in the numerator, our conditional probability is for sufficiently
large $n$ as close as we want to $1$.

\section{Weak perturbations of Gaussian measures:\\
a non Gaussian fixed point}
Starting at the end of the seventies the RG has become a very
important and effective tool for proving rigorous results in
statistical mechanics and Euclidean quantum field theory.
An impressive amount of work has been done and it is not
possible to give even a schematic account of it \cite{GA}. Many different
versions of the RG idea have been used, each author or group
of authors following his own linguistic propensities. Probability
theory is always in the background and we want to try to
recover some conceptual feature common to all of them.
As in the previous part of this review limit theorems will be
a relevant reference. However the limit theorems to be
considered are of a different kind, they are those which in probability
are called non classical and are related to the following problem. 

Given a probability distribution $P$ and an integer $n$, can one
consider it as resulting from the composition (convolution) of $n$
distributions $P_k,~ k=1,2,\ldots,n$? In other words can one consider
the random variable described by $P$ as the sum of $n$ independent
random variables? In formulas
\begin{equation}
P=P_1\star P_2\star \ldots \star P_n,
\label{CONV}
\end{equation}
where $\star$ means convolution. 
It is clear, for example,
that a Gaussian distribution of variance $\sigma^2$ can be thought as
the composition of any number $n$ of Gaussians with
variances $\sigma_i^2$ provided $\sum_1^n \sigma_i^2=\sigma^2$. 
The problem arises naturally of investigating under what
conditions convolutions like the right hand side of (\ref{CONV})
converge to a regular distribution as $n\rightarrow\infty$.
The right hand side of (\ref{CONV}) can  be considered as a recurrence relation
\begin{equation}
{\hat P}_{n+1}={\hat P}_n\star P_n,
\label{REC}  
\end{equation}
where ${\hat P}_n=P_1\star P_2\star \ldots \star P_{n-1}$.
Comparing (\ref{R}) with (\ref{CONV}) we see that while in the
case of limit theorems for independent identically distributed random variables
we have a natural fixed point problem, this is not in general the case for
non identically distributed  variables.  
As we shall see below, the RG approach to Euclidean
field theory and the statistical mechanics of the critical point has
led to formulations which have analogies with these
problems.
In fact, infinite dimensional equations
structurally similar to (\ref{REC}) are constructed which can be
transformed into equations admitting fixed points after a rescaling. 

In the following exposition we shall follow
the recent article by Brydges, Dimock and Hurd \cite{BDH}.  The goal of
these authors is the construction
of a quantum field theory in ${\bf {R}}^4$ with non trivial scaling
behaviour at long distances, that is in the infrared region, determined
by a non Gaussian fixed point of an appropriate RG
transformation. The starting point is a $\phi^4$ theory in finite
volume regularized at small distances to eliminate ultraviolet
singularities. This model is believed to have a non Gaussian 
fixed point in  $4-\epsilon$ dimensions and to simulate such a
situation in $4$ dimensions the authors introduce a special covariance
for the Gaussian part of the measure.
The first step consists in the construction of a covariance $v(x-y)$
which behaves at large distances like $(-\Delta)^{-1-\epsilon /2}$.
This means that it scales like $|x|^{-2+\epsilon}$ for large $|x|$.
Their choice is
\begin{equation}
v(x-y)=\int_1^\infty d\alpha \alpha^{\epsilon/2-2}e^{-|x-y|^2/4\alpha}.
\end{equation}
Such a covariance can be decomposed in the following way
\begin{equation}
v(x-y)=\sum_{j=0}^\infty L^{-(2-\epsilon)j}C(L^{-j}(x-y)),
\end{equation}
where $L>1$ is a scaling factor and
\begin{equation}
C(x)=\int_1^{L^2} d\alpha \alpha^{\epsilon/2-2}e^{-|x|^2/4\alpha}.
\end{equation}
Each term in the expansion can be interpreted as the covariance
of a rescaled field
\begin{equation}
\phi_{L^{-j}}(x)=L^{-(2-\epsilon)j/2}\phi(L^{-j}x)
\end{equation}
which has reduced fluctuations and varies over larger distances.
The aim is to study  the measure
\begin{equation} 
d\mu_\Lambda=Z^{-1}e^{-V_\Lambda(\phi)}d\mu_v
\end{equation}
where
\begin{equation}
V_\Lambda(\phi)=\lambda\int_\Lambda {:\phi^4:}_v + 
\zeta \int_\Lambda {:(\partial \phi)^2:}_v + 
\mu \int_\Lambda {:\phi^2:}_v
\end{equation}
when $\Lambda$ tends to ${\bf R}^4$. The double dots indicate
the Wick polynomials with respect to the covariance $v$.

Take for $\Lambda$ a large cube of side $L^N$ so that the measure is
well defined. We want to calculate
\begin{equation}
(\mu_v\star e^{-V})(\phi)=(\mu_{{\hat C}_N}\star\mu_{{\hat
C}_{N-1}}\star ....\star\mu_{{\hat C}_0}\star e^{-V})(\phi)
\end{equation}
having used the above decomposition of the covariance with
\begin{equation}
{\hat C}_j (x)=L^{-(2-\epsilon)j}C(L^{-j}(x-y))
\end{equation}
Actually in finite
volume we should specify some boundary conditions but we
shall ignore this aspect. Next by defining
\begin{equation}
{\hat Z}_j (\phi)= (\mu_{{\hat C}_{j-1}}\star \ldots \star\mu_{{\hat
C}_0}\star e^{-V})(\phi)
\end{equation}
we find the recursion relation
\begin{equation}
{\hat Z}_{j+1} (\phi)=(\mu_{{\hat C}_j}\star {\hat Z}_j) (\phi).
\end{equation}
In this way the calculation is performed by successive integrations
over variables which exhibit decreasing fluctuations. 
This is not yet our RG equation because as $j\rightarrow\infty$, $\mu_{{\hat C}_j}$ 
becomes a singular distribution and we do not obtain a fixed point equation.
However by introducing the rescaled fields 
$\phi_{L^{-j}}(x)=L^{-(2-\epsilon)j/2}\phi(L^{-j}x)$ and the rescaled
$Z$'s 
\begin{equation}
Z_j(\phi)={\hat Z}_j(\phi_{L^{-j}})
\end{equation}
the recursion becomes
\begin{equation}
 Z_{j+1}(\phi)=(\mu_{\hat C} \star  Z_j) (\phi_{L^{-1}})
\label{RC}
\end{equation}
with initial condition $Z_0=e^{-V}$.
We emphasize that the last step is possible due to the special structure of the measures
$\mu_{{\hat C}_j}$. 
It is now meaningful to look for the fixed points of (\ref{RC}).
Brydges, Dimock and Hurd have proved that in $d=4$ for $\epsilon$ small there exists
a non Gaussian  
fixed point of (\ref{RC}) characterized by a value $\hat {\lambda}(\epsilon, L)$ of the
coupling $\lambda$ and that for certain values $\mu({\lambda}), \zeta({\lambda})$
the iteration of (\ref{RC}) with initial condition $e^{-V}$ converges to this fixed point. 
Technically the proof
is very complicated and its description is beyond the aims of this review. A very
good exposition with some simplifications of the techniques employed can be found
also  in \cite{MS}. 

\section{Concluding Remarks}
The question we want to consider is the following: which are the 
benefits for our understanding of critical phenomena 
and more generally of statistical physics deriving from the use
of probabilistic language? Feynman thought that it is worth to
spend one's time formulating a theory in every physical and
mathematical way possible. In our case there is an intuition associated
with probabilistic reasoning that is foreign to the usual 
formalisms of statistical mechanics based on correlation
functions and equations connecting them.

Apart from this general remark we must consider that the
rigorous results obtained so far in RG theory
have been strongly influenced by
the probabilistic language as this appears the most natural
for the mathematical study of statistical mechanics
and Euclidean field theory when a functional integral approach
is used.
New technical ideas however are
needed to deal with  concrete problems like
calculating the critical indices of the $3$-dimensional Ising model
or establishing in a conclusive way whether the field
theory $\phi^4_4$ is ultraviolet non trivial.   

The formal apparatus of RG has been easily extended
to the analysis of fermionic systems when these
are described by a Grassmaniann functional integral \cite{BG},
that is by the analog of a Gibbs distribution over
anticommuting variables. In this case the convergence
of perturbation theory plays a major role on the way to rigorous results. 
Recently, it has been possible 
to give a true probabilistic expression to general Grassmaniann integrals
in terms of discrete jump processes (Poisson processes)
\cite{DJS}, \cite{BPDJ} so that classical probability
may become   a main tool also in the study of fermionic 
systems especially in view of developing non perturbative
methods. For an early example of connection between
anticommutative calculus and probability see \cite{BMM}.

In a wider perspective one may remark that  
the theory of Gibbs distributions 
is becoming instrumental also
in various sectors of mathematical statistics, for example
in image reconstruction, and critical situations appear also
in this domain. 
Transfer of ideas from statistical mechanics to stochastic analysis
is currently an ongoing process which shows the relevance
of a language capable of unifying different areas of research.  
Probability theory for a long time
has not been included among the basic mathematical tools
of a physics curriculum but the situation is slowly changing
and hopefully this will help cross fertilization 
among different disciplines.

%\appendix*{Appendix}

%%%%%%%%%%%%%%%%%%%%%%%%%%%%%%%%% ACKNOWLEDGMENTS
\acknowledgments 
This paper is an expanded version of a talk given at the meeting RG 2000,  Taxco,
Mexico,  January 1999. It is a pleasure to thank C. Stephens and D. O' Connor
for their kind invitation.

%
%%%%%%%%%%%%%%%%%%%%%%%%%%%%%%%%% REFERENCES LIST
%

\end{document}